%% file: LightQuarkStructure.tex
\RequirePackage{lineno}
\documentclass[aps,prl,superscriptaddress,groupedaddress,preprintnumbers]{revtex4}  
\usepackage{graphicx}  
\usepackage{dcolumn}   
\usepackage{bm}        
\usepackage{amssymb}   
\usepackage{multirow}
\usepackage{epsfig}
\usepackage{amsmath}
\usepackage[colorlinks, linkcolor=blue]{hyperref}
\usepackage{ulem}
\usepackage{appendix}
\usepackage{lineno}
\usepackage{setspace}

\newcommand{\Dzero}{D0~}
\newcommand{\chisquare}{\mathcal{X}^2}
\newcommand{\chEf}{$C_H(E,f)$ }
\newcommand{\xsim}{$x \sim 0.1$}
\newcommand{\resbos}{\textsc{ResBos}}
\newcommand{\mcfm}{\textsc{MCFM}}
\newcommand{\epump}{\textsc{ePump}}

\begin{document}
\lefthyphenmin=2
\righthyphenmin=3


\title{Impact of new measurements of light quarks at hadron colliders}
\input author_list.tex

\begin{abstract} 
	{\bf Abstract}: Recently a series of new measurements with both the neutral and charge current Drell--Yan processes have been performed at hadron colliders, showing deviations from the predictions of the current parton distribution functions (PDFs). In this article, the impact of these new measurements is studied by using their results to update the PDFs. 
Although these new measurements correspond to different boson propagators and colliding energies, they are found to have a similar impact to the light quark parton distributions with the momentum fraction $x$ around 0.1. 
It manifests that the deviations are consistent with each other and favor a larger valence $d_v/u_v$ ratio than the modern PDF predictions. Further study indicates that such tension arises dominantly from the deep inelastic scattering measurements of NMC and the fixed target experiments of NuSea, both of which play pivotal roles in detecting the relative $u$ and $d$ type quark contributions for modern PDFs. According to the conclusions of the impact study, it would be essential to include these new measurements into the complete PDF global analysis in the future.

\vspace{0.7em}
\end{abstract}

\maketitle

\section{Introduction}
\label{sec:intro}

The Drell--Yan processes~\cite{dy}, encompassing both neutral current $hh(q_i\bar{q_i})\rightarrow Z/\gamma^* \rightarrow \ell^+\ell^-$, and charged current $hh(q_i\bar{q_j})\rightarrow W^{\pm} \rightarrow \ell^{\pm} \nu$ production, constitute one of the most crucial inputs for proton parton distribution function (PDF) analyses. Within the framework of Quantum Chromodynamics (QCD), inclusive Drell--Yan process is rigorously factorized~\cite{factorization} into a short-distance hard-scattering part calculable in perturbation theory and a universal long-distance non-perturbative part incorporated in the PDFs. This established factorization enables the Drell--Yan processes to serve as model-independent probes for PDF determination. At hadron colliders, the productions of Drell--Yan processes are primarily initiated by $u$ ($\bar{u}$) and $d$ ($\bar{d}$) quarks, which play a fundamental role in the PDF studies. 
The colliding energy and kinematic features renders the Drell--Yan process sensitive to a wide range of the momentum fraction $x$ of the initial-state quarks, especially the region $x \sim$ 0.1, where the so-called ``valence'' quarks $u_v$ and $d_v$ are strongly dominant. 
Furthermore, the Drell--Yan processes exhibit exceptional precision in both theoretical predictions and experimental measurements. Theoretically, the inclusive Drell--Yan production allows the resummation and fixed-order perturbative QCD computations achieving percent-level accuracy; experimentally, the final-state leptons originating from $W/Z$ boson decays are detected with very high efficiencies and small uncertainties. Together with very small background contaminations, this enables experimental uncertainties well controlled at also percent level.   
Under these conditions, once new Drell--Yan measurements with high precision become available, it's crucial to examine their impacts to light quark PDFs. When deviations are observed between data and predictions, it may imply potential limitations in current PDF models or QCD calculations. Such discrepancies offer unique opportunities to refine our understanding of proton structure. 

Some recent Drell--Yan measurements, including the proton structure parameter $R$ (closely approximating  $d_v/u_v$)~\cite{AFBFactorization} extracted from the forward–backward asymmetry ($A_{FB}$) in $Z/\gamma^* \to \ell^+\ell^-$ by the \Dzero~\cite{D08ifb} and CMS~\cite{CMS} collaborations, as well as the $W$ charge asymmetry measurement at high transverse mass ($m_T$) region  by ATLAS~\cite{wasy}, collectively reveal significant deviations from current PDF predictions in the parton momentum fraction range \xsim. 

These new datasets exhibit distinctive characteristics. First, they can provide flavor-sensitive information about the relative contributions of $u$ and $d$ quarks. Although various Drell--Yan measurements~\cite{hera, d01, cdf1, lhcb1, lhcb2, atlas1, atlas2, cdf2, d02, cms1, atlas3, d03, cms2, bcdms1, bcdms2, nmc, dimuon, nusea, nusea2} have been extensively incorporated into modern PDFs~\cite{CT18NNLO, MSHT20, NNPDF4}, direct experimental determination of the relative $u$ and $d$ quark distributions, such as the light quark ratio $d/u$, remains very rare. This difficulty arises because the Drell--Yan production occurs through both $u\bar{u}$ and  $d\bar{d}$ initial states with comparable cross sections and identical final-state particles. Consequently, the $u$($\bar{u}$) and $d$($\bar{d}$) contributions are always mixed together and experimentally indistinguishable, which makes the $u$ and $d$ quarks determination heavily depend on the choice of non-perturbative formalism in the global analysis. Second, with higher collider energies, the $s$-, $c$-, and $b$-type quark contributions become more significant,  thereby complicate the dependence on non-perturbative assumptions in the PDF global fits. Fortunately, observables from these new datasets are deliberately designed to contain no or insignificant contributions from these heavy quarks.  Due to their significant deviations from PDF predictions, the distinctive features make the impact study of these new datasets on PDFs particular crucial.

The study in this paper verifies that deviations from all the three datasets are mutually consistent, which may indicate a coherent tendency in the $u$ and $d$ quark behaviors with respect to the current PDFs. Details are organized as follows: Section~\ref{sec:data} presents a short review of the three new Drell--Yan measurements; Section~\ref{sec:corr} studies the correlations between these new measurements and the PDFs using the cosine of the Pearson correlation angle; Section~\ref{sec:tension} examines the impact of these new measurements by updating the PDFs with their results; Section~\ref{sec:sum} presents a summary.

\section{Review of the new Drell--Yan measurements}
\label{sec:data}

For the neutral current Drell--Yan measurement, a recent method proposed in Reference~\cite{AFBFactorization} enables a refined flavor decomposition through the forward-backward asymmetry ($A_{FB}$) spectrum in the $Z/\gamma^* \rightarrow \ell^+\ell^-$ Drell--Yan process. The $u$ ($\bar{u}$) and $d$ ($\bar{d}$) quark information inside $A_{FB}$ observable can be factorized into a well-defined structure parameter, $R$~\cite{CMS}, which serves as a novel experimental observable and reflects unique information of the relative difference between $u$ and $d$ quarks. 

At a $p\bar{p}$ collider such as the Tevatron and a $pp$ collider such as the LHC, the observable $R$ can be approximately expressed as
\begin{equation}
	R_{p\bar{p}} \propto \frac{d(x_1)\,d(x_2) - \bar{d}(x_1)\bar{d}(x_2)}{u(x_1)\,u(x_2) - \bar{u}(x_1)\bar{u}(x_2)}~, \quad
	R_{pp} \propto \frac{d(x_1)\,\bar{d}(x_2) - \bar{d}(x_1)d(x_2)}{u(x_1)\,\bar{u}(x_2) - \bar{u}(x_1)u(x_2)}~.
    \label{equ:apr1}
\end{equation}
In the above definitions, $x_{1,2}$ are the momentum fractions of the two initial-state partons given by 
$x_{1,2} = \frac{\sqrt{M^2 + Q_T^2}}{\sqrt{s}} e^{\pm Y}$, where $M$, $Y$ and $Q_T$ are the invariant mass, rapidity and transverse momentum of the dilepton system in Drell--Yan productions, and $\sqrt{s}$ is the collision center-of-mass energy. The convention $x_1 > x_2$ is followed, reflecting the typical kinematic configuration 
at hadron colliders, where one parton carries a substantially larger momentum fraction than the other due to the boost of the dilepton system. The contributions from $s$-, $c$- and $b$-type quarks are nearly perfectly cancelled due to $q(x,Q) = \bar{q}(x,Q)$, for $q = s$, $c$, and $b$  up to the next-to-leading order (NLO) in QCD interactions, assuming initial equality at $Q_0$ around 1 GeV.
Given the approximation that the light-quark distributions are approximately flavor-symmetric at low $x$~\cite{CT18NNLO} (i.e.,$u(x_2) \approx \bar{u}(x_2)\approx d(x_2) \approx \bar{d}(x_2)$) for $x_2 \lesssim 5\times10^{-3}$), one can extract a ratio at both $pp$ collider and $p\bar{p}$ collider:
\begin{equation}
	R \propto \frac{d(x_1) - \bar{d}(x_1)}{u(x_1) - \bar{u}(x_1)} = \frac{d_v(x_1)}{u_v(x_1)},
	\label{equ:apr2}
\end{equation}
which provides a clean experimental probe of the valence quark ratio $d_v/u_v$ in the \xsim~region.

\begin{figure}[hbtp!]
	\centering
	\includegraphics[width=0.32\linewidth]{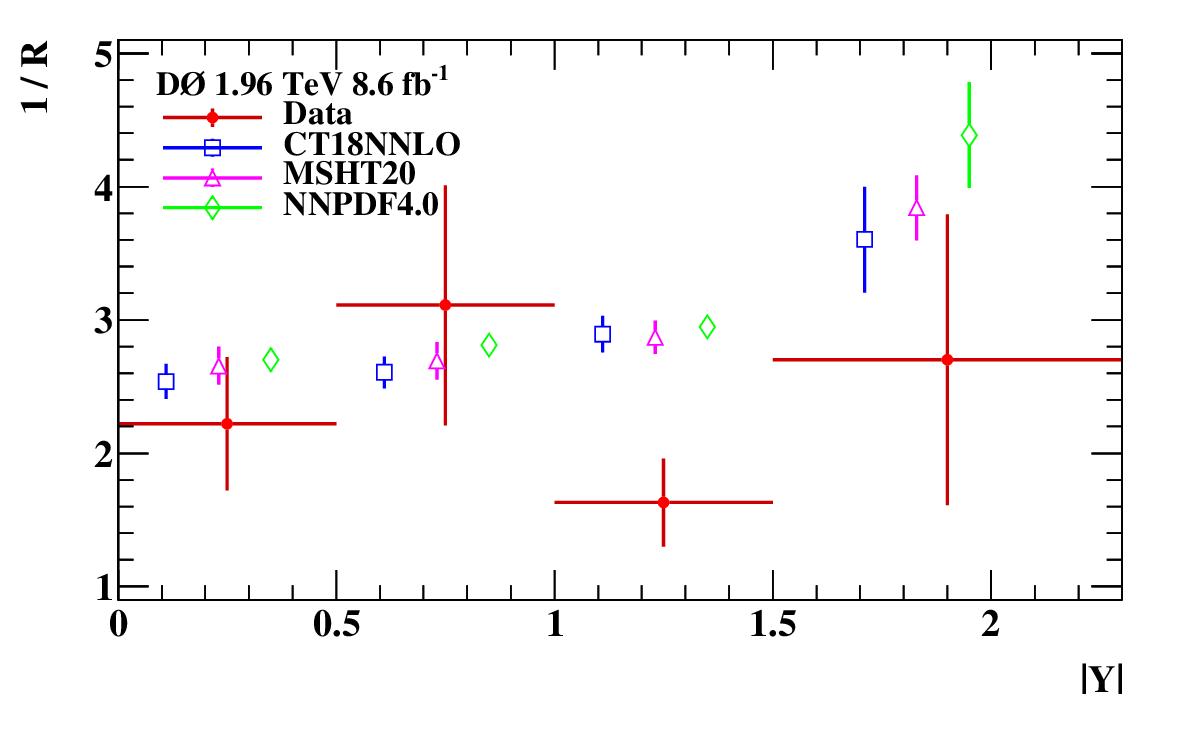}
	\includegraphics[width=0.32\linewidth]{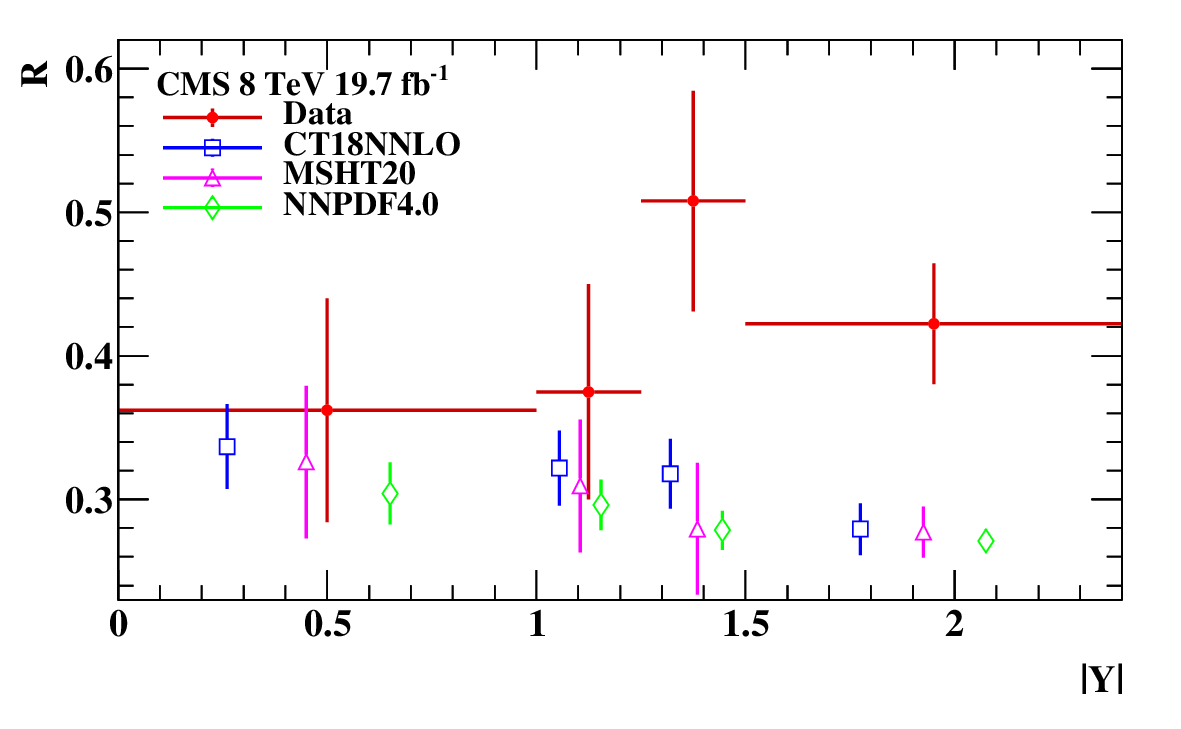}
	\includegraphics[width=0.32\linewidth]{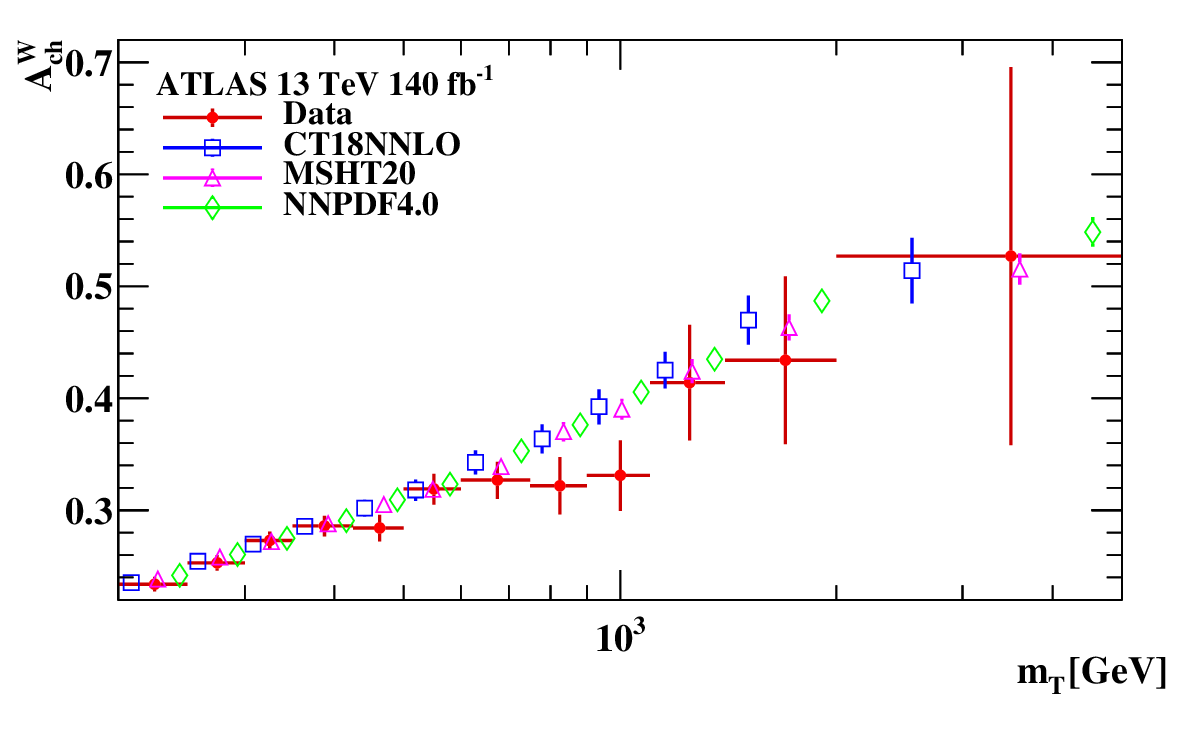}
	\caption{Comparison between experimental measurements and theoretical predictions obtained using the modern PDF sets CT18NNLO~\cite{CT18NNLO}, MSHT20~\cite{MSHT20}, and NNPDF4.0~\cite{NNPDF4}, for observables sensitive to the flavor composition of the proton in hadron collider environments. The left panel shows the \Dzero measurement of the inverse $1/R$ in the Drell--Yan process at $\sqrt{s} = 1.96$ TeV, while the middle panel presents the CMS measurement of $R$ at $\sqrt{s} = 8$ TeV in a similar process. The right panel shows the ATLAS measurement of the $W$ boson charge asymmetry $A^W_\text{ch}$ at $\sqrt{s}$ = 13 TeV. According to Equation~\ref{equ:apr2}, $R$ is defined as a function of $x_1$, with $x_1 \sim 0.1$ being a valid approximation only in the rapidity ranges $1<|Y|<1.5$ for the \Dzero data and $1.25<|Y|<2.4$ for the CMS data. The complete set of rapidity bins in the left (middle) panel covers a wider range of $0.05 < x_1 < 0.4$ ($0.005 < x_1 < 0.1$) values.} 
	\label{fig:AllData}
\end{figure}

These analyses have been performed: one with the $R$ parameter extracted using 1.96 TeV $p\bar{p}$ collision data collected with the D0 detector at the Tevatron~\cite{D08ifb}, and the other with the $R$ parameter extracted using 8 TeV $pp$ collision data collected with the CMS detector at the LHC~\cite{CMS}. Both results indicate an enhancement of the $d_v/u_v$ ratio in the  $x$ region around $0.1$ relative to the predictions from current PDF sets.
These measurements are expected to provide unique information in the PDF analysis:
the \Dzero measurement extracted the structure parameter $R$ differentially in $Z$ boson rapidity regions, focusing in particular on the interval $1 < |Y| < 1.5$, which corresponds to $x_1 \sim 0.1$ and $x_2 \sim 0.01$. In this kinematic region, the measured $R$ values were found to be significantly higher than the predictions of the three modern PDFs (CTEQ~\cite{CT18NNLO}, MSHT~\cite{MSHT20} and NNPDF~\cite{NNPDF4}) by more than 3.5 standard deviations as illustrated in Figure~\ref{fig:AllData} (left panel). For the same observable extraction using CMS data in the forward rapidity intervals ($1.25 < |Y| < 2.4$), corresponding to $x_1 \sim 0.05$-0.1 and $x_2 \sim 0.002$, the measured $R$ values were also found to be significantly larger than the predictions of the three PDFs, again implying an ehanced $d_v/u_v$ ratio in \xsim~region (see Figure~\ref{fig:AllData}, middle panel). Given the absence of direct constraints on the relative composition of valence quarks in current global PDF fits, such deviations are acceptable. 

For the charged Drell--Yan process, ATLAS collaboration performed a $W$ charge asymmetry ($A^W_\text{ch}$) measurement in the high transverse mass region ($m_T$ up to 5 TeV) at $\sqrt{s} = 13$ TeV~\cite{wasy}.
The $W$ charge asymmetry is a classic experimental observable sensitive to PDF defined as  $A^W_\text{ch} = (\sigma_{W^+} - \sigma_{W^-})/(\sigma_{W^+} + \sigma_{W^-})$, where $\sigma$ is the measured cross sections. Compared with the traditional $W$ charge asymmetry measurements, the high transverse mass requirement allows access to higher values of $x$, such as \xsim. 
In this $x$ region, the initial $u$ and $d$ quarks lie very close to the valence peak distributions, whereas $s$($\bar{s}$), $c$($\bar{c}$) and $b$($\bar{b}$) quarks reside in their tail parts. Consequently, the processes are overwhelmingly dominated by $u$ and $d$ quarks. This configuration serves as an independent probe for studying the relative $d/u$ quark composition.
Notably, a deficit is observed in the ATLAS results relative to the PDF prediction around $m^W_T \sim 1$ TeV, suggesting that the relative structure of $u$ and $d$ quarks may differ from current PDF expectations (see Figure~\ref{fig:AllData}, right panel). The corresponding $x_1$ values probed in this high-$m^W_T$ region lie between 0.02 and 0.3, overlapping with the $x_1$ range probed by the $R$ measurements from both the D0 and CMS experiments.

Given that both observables, the high transverse mass region $W$ charge asymmetry and the $R$ from $A_{FB}$, can provide information on the relative $u$ and $d$ quark contributions and show deviations from current PDF predictions, especially considering that these measurements are performed at different colliders with different center-of-mass energies, involving different processes and distinct production mechanisms for accessing the \xsim~region, their mutual consistency is particularly interesting. As there are two quarks in the initial states, $q(x_1)\bar{q}^{(\prime)}(x_2)$ and $\bar{q}^{(\prime)}(x_1)q(x_2)$ are alway mixed together, it is necessary to quantitatively investigate the correlations among these datasets and their collective impact on the $u$ and $d$ quark PDFs of the proton.

\section{Dataset PDF Correlation Patterns and Implications}
\label{sec:corr}

To examine the relative correlation and mutual compatibility of the experimental datasets, a commonly used approach is to compute the cosine of the Pearson correlation angle $C_H(X,Y)$~\cite{cosh} between two quantities $X$ and $Y$. In the multidimensional parameter space of the Hessian PDF framework~\cite{epump, hessian1, hessian2}, this evaluation is performed by computing the variations of $X$ and $Y$ under unit displacements along each eigenvector direction $k$ of the PDF error set. Specifically, $X_{\pm k} \equiv X(0, ..., \pm1, ..., 0)$ and $Y_{\pm k}$ denote the values of $X$ and $Y$ evaluated at the $\pm1\sigma$ positions along the $k$-th eigenvector direction. The uncertainties $\delta_H X$ and $\delta_H Y$ are defined as: 
$\delta_H X = \frac{1}{2}\sqrt{\sum_{k=1}^{D}\left(X_{+k}-X_{-k}\right)^2}$, 
$\delta_H Y = \frac{1}{2}\sqrt{\sum_{k=1}^{D}\left(Y_{+k}-Y_{-k}\right)^2}$. 
The correlation angle is then computed as:
\begin{equation}
	C_H(X, Y) = \frac{1}{4\delta_H X \delta_H Y} \sum_{k=1}^{D} (X_{+k}-X_{-k})(Y_{+k}-Y_{-k}) = \frac{\sum_{k=1}^{D} (X_{+k}-X_{-k})(Y_{+k}-Y_{-k})}{\sqrt{\sum_{k=1}^{D} (X_{+k}-X_{-k})^2} \sqrt{\sum_{k=1}^{D}(Y_{+k}-Y_{-k})^2}}
	\label{equ:chXY}
\end{equation}
Under this convention, the cosine of the correlation angle between the minimal fit $\chisquare$ for dataset $E$ and the parton distribution function $f$ is defined as:
\begin{equation}
	C_H(E, f) = \frac{\sum_{k=1}^{D} ({\chisquare_E}_{(+k)}-{\chisquare_E}_{(-k)})(f_{+k}-f_{-k})}{\sqrt{\sum_{k=1}^{D} ({\chisquare_E}_{(+k)}-{\chisquare_E}_{(-k)})^2} \sqrt{\sum_{k=1}^{D}(f_{+k}-f_{-k})^2}}
\end{equation}
Thus, a positive \chEf value generally implies that the direction minimizing $\chisquare_E$ simultaneously reduces the value of PDF $f$. Conversely, a negative \chEf indicates that dataset $E$ preferentially enhances $f$.

Since $\chisquare_{E}$ depends on the theoretical predictions for the observables in dataset $E$, the \resbos~generator~\cite{resbos} is employed for neutral Drell--Yan process predictions, and \mcfm~\cite{mcfm} is employed for charged Drell--Yan process predictions. \resbos~provides approximate \textsc{NNLO} QCD + \textsc{N${}^{3}$LL} resummation and electroweak corrections via the effective Born approximation~\cite{born}. 
The \mcfm~program provides NNLO QCD predictions for W-boson production, incorporating NLO electroweak corrections. 

\begin{figure}[htb!]
	\centering
	\includegraphics[width=0.32\linewidth]{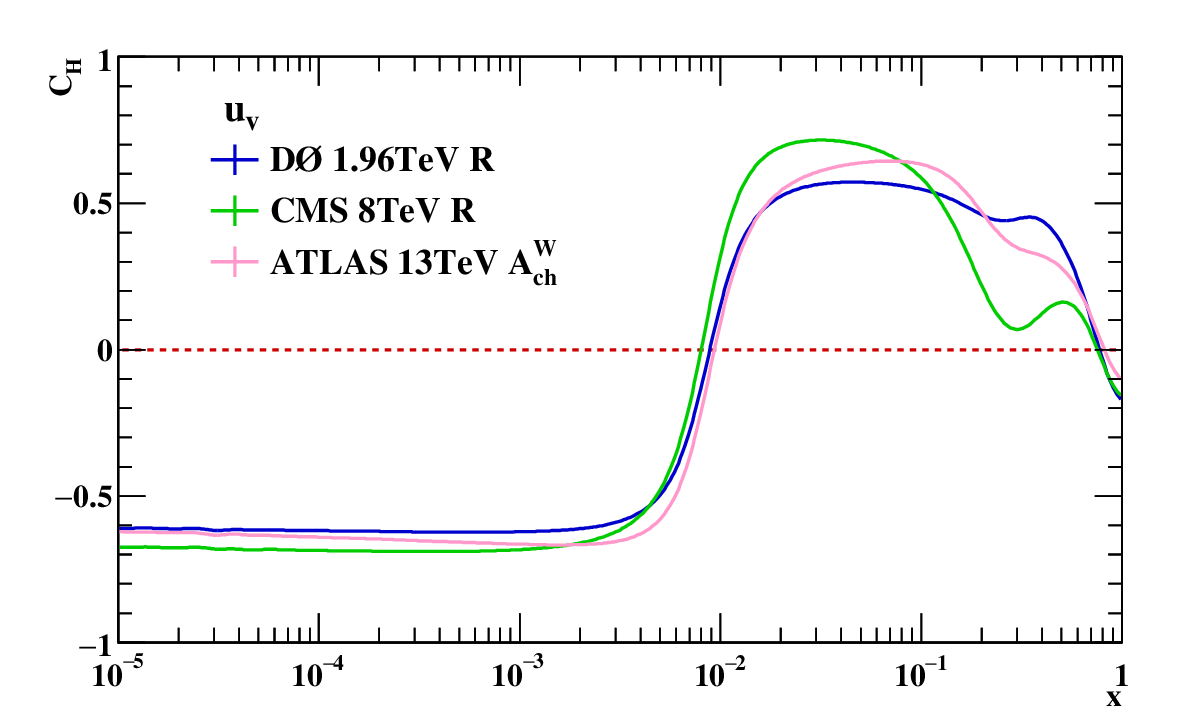}
	\includegraphics[width=0.32\linewidth]{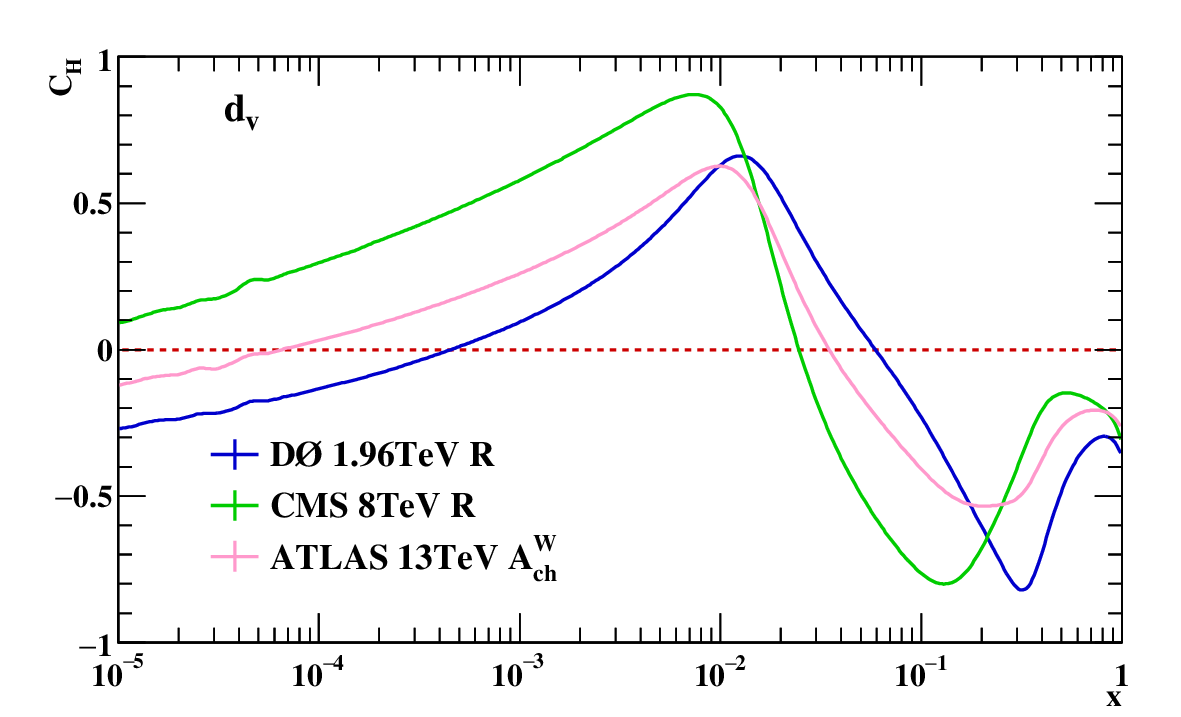}
	\includegraphics[width=0.32\linewidth]{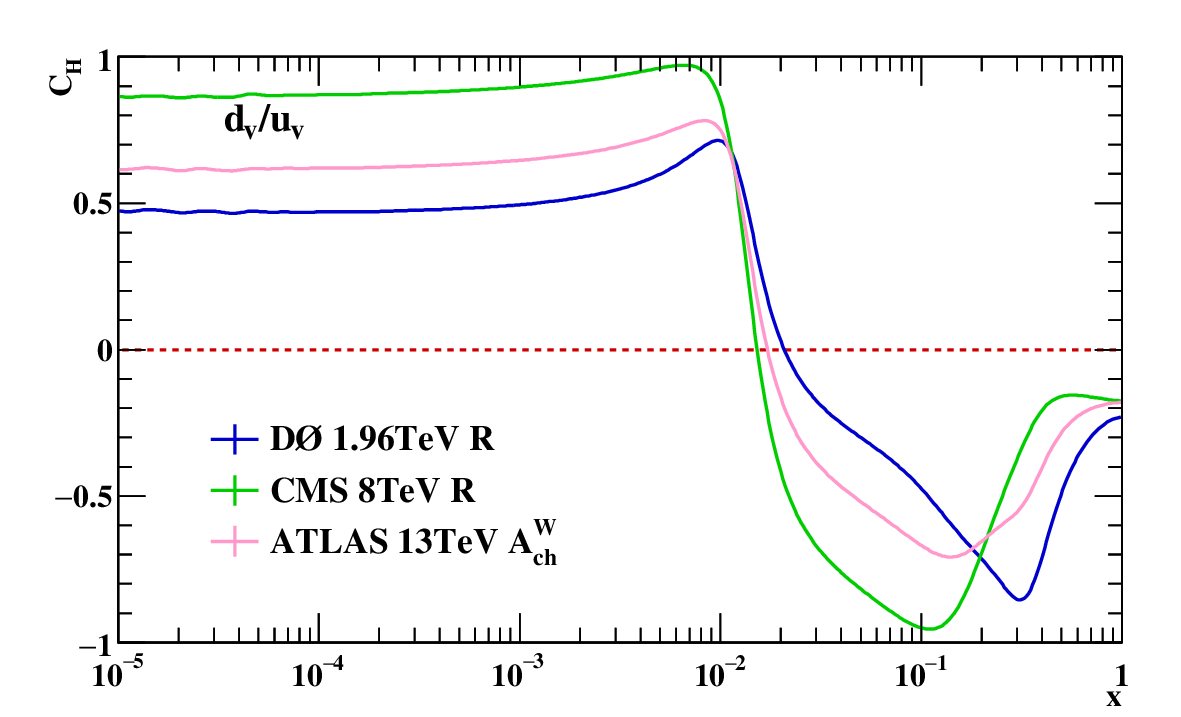}\\
	\includegraphics[width=0.32\linewidth]{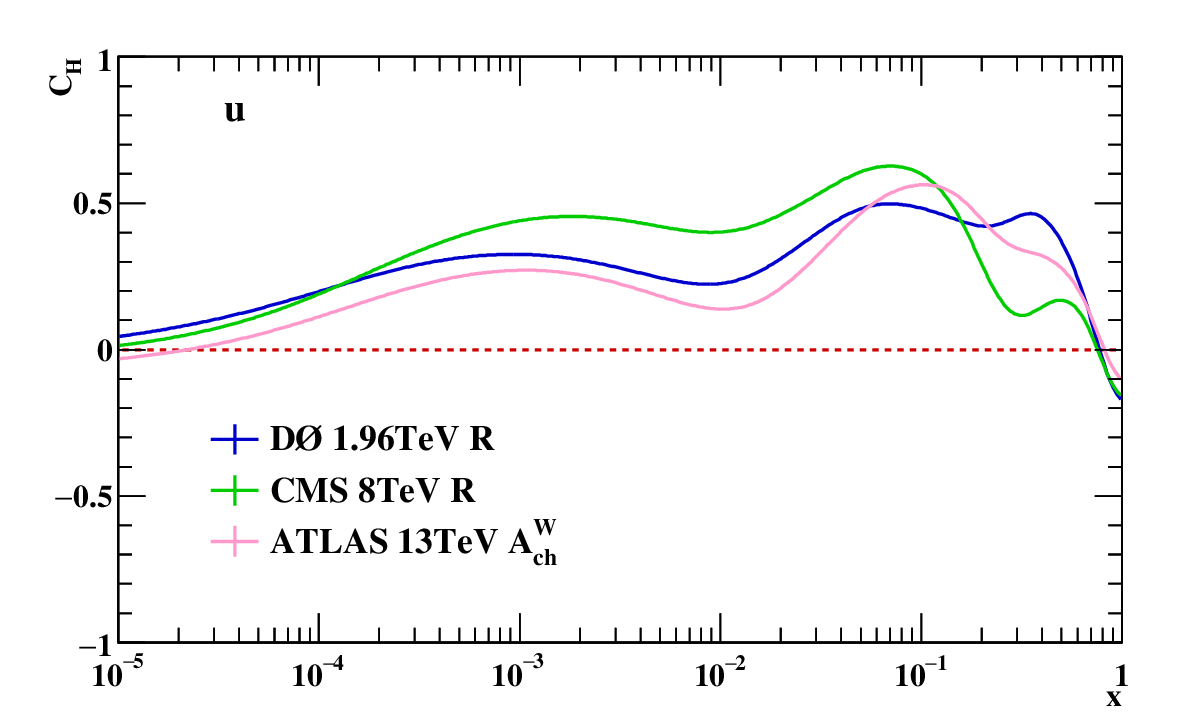}
	\includegraphics[width=0.32\linewidth]{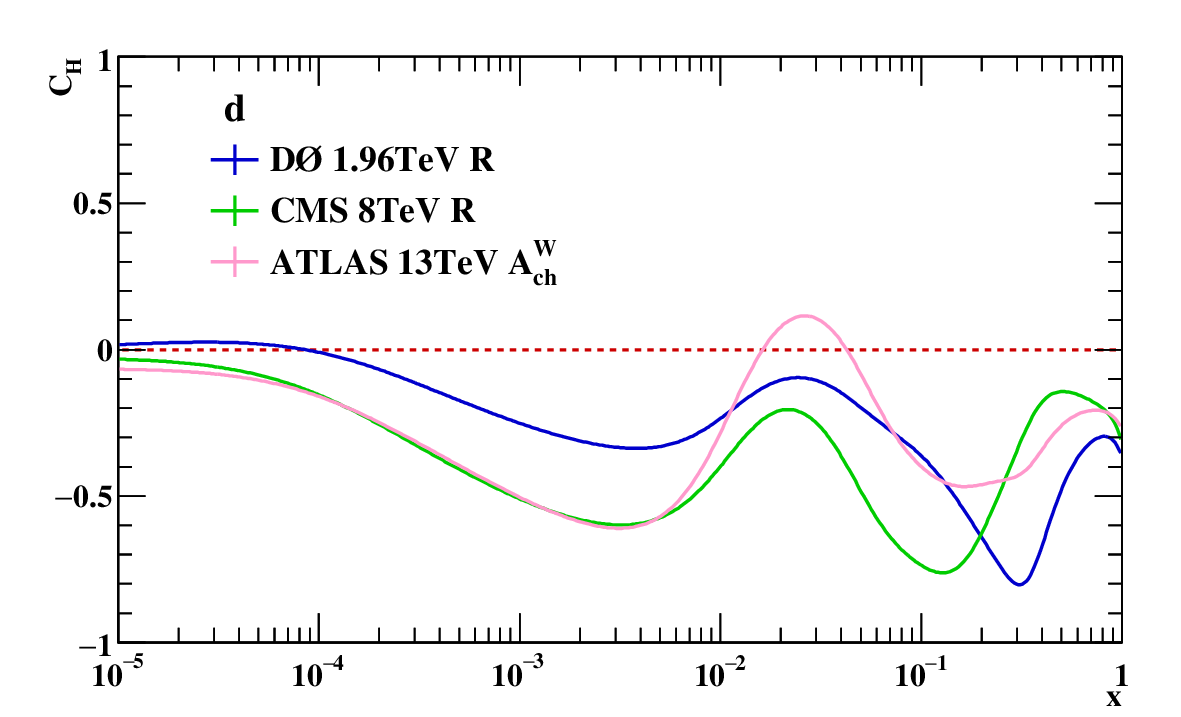}
	\includegraphics[width=0.32\linewidth]{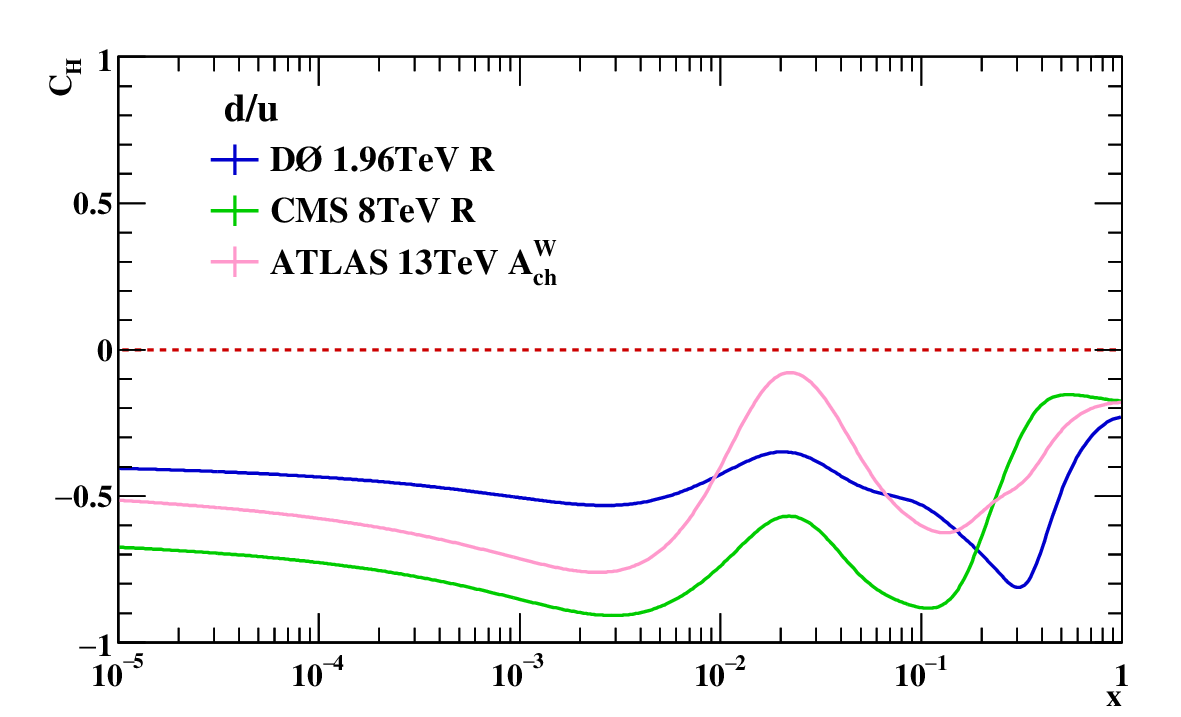}\\
	\includegraphics[width=0.32\linewidth]{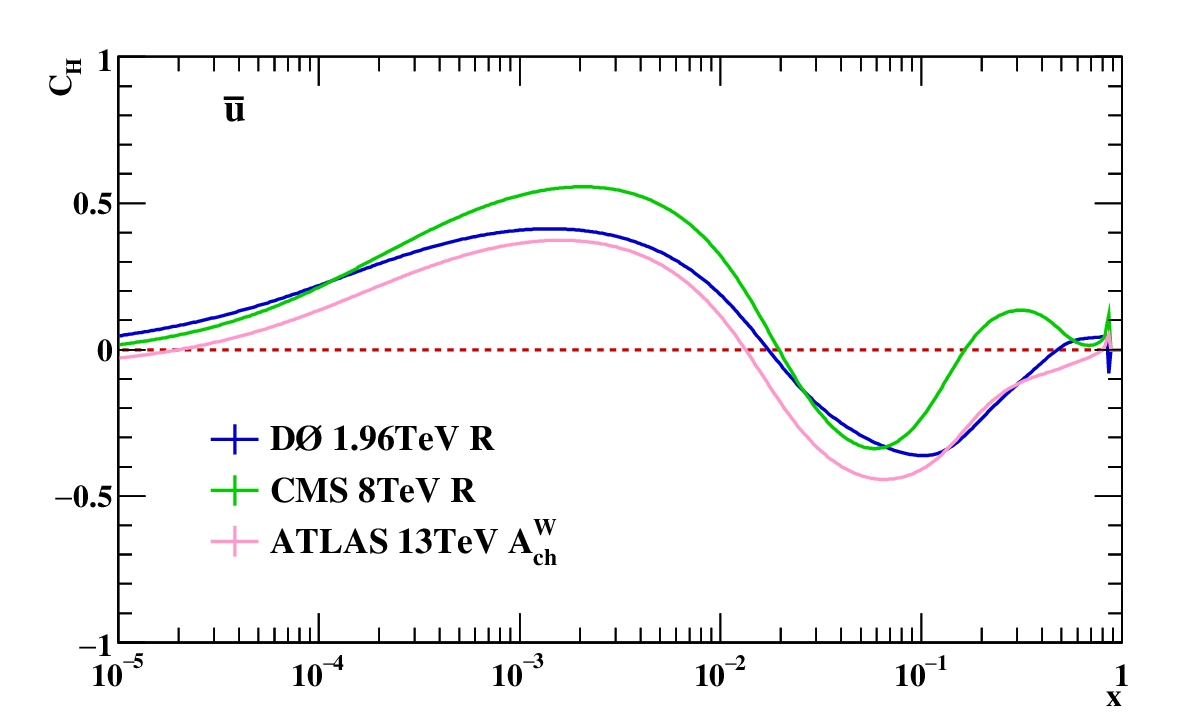}
	\includegraphics[width=0.32\linewidth]{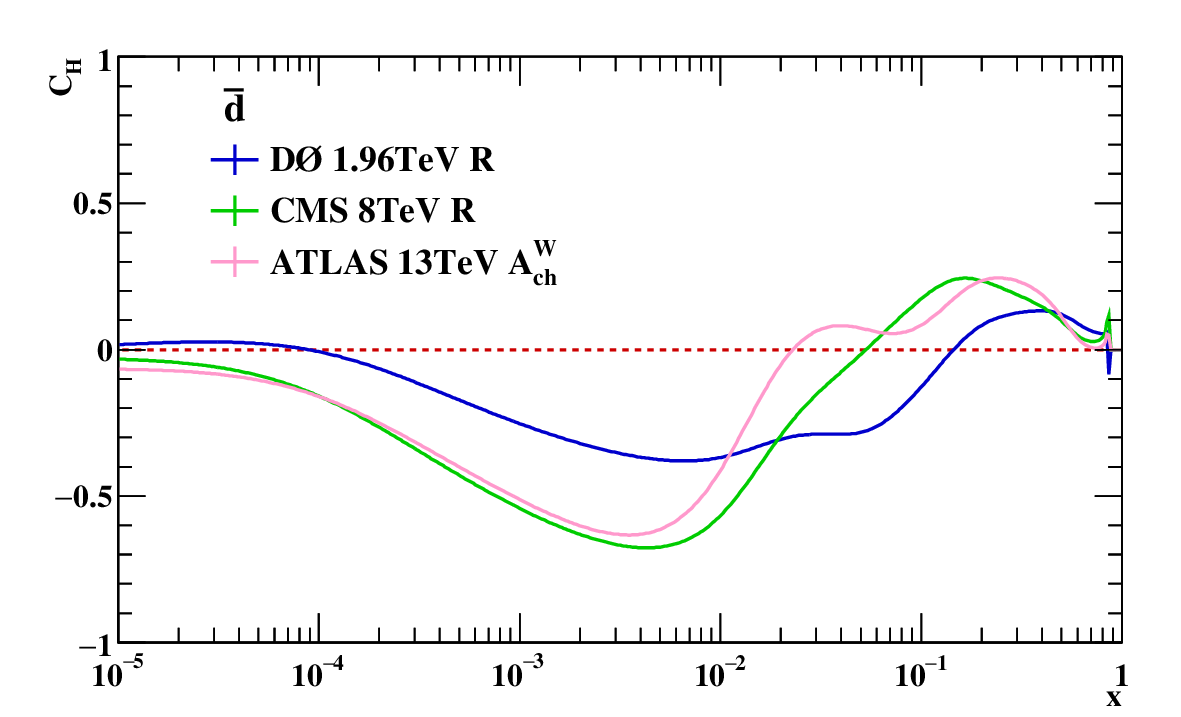}
	\includegraphics[width=0.32\linewidth]{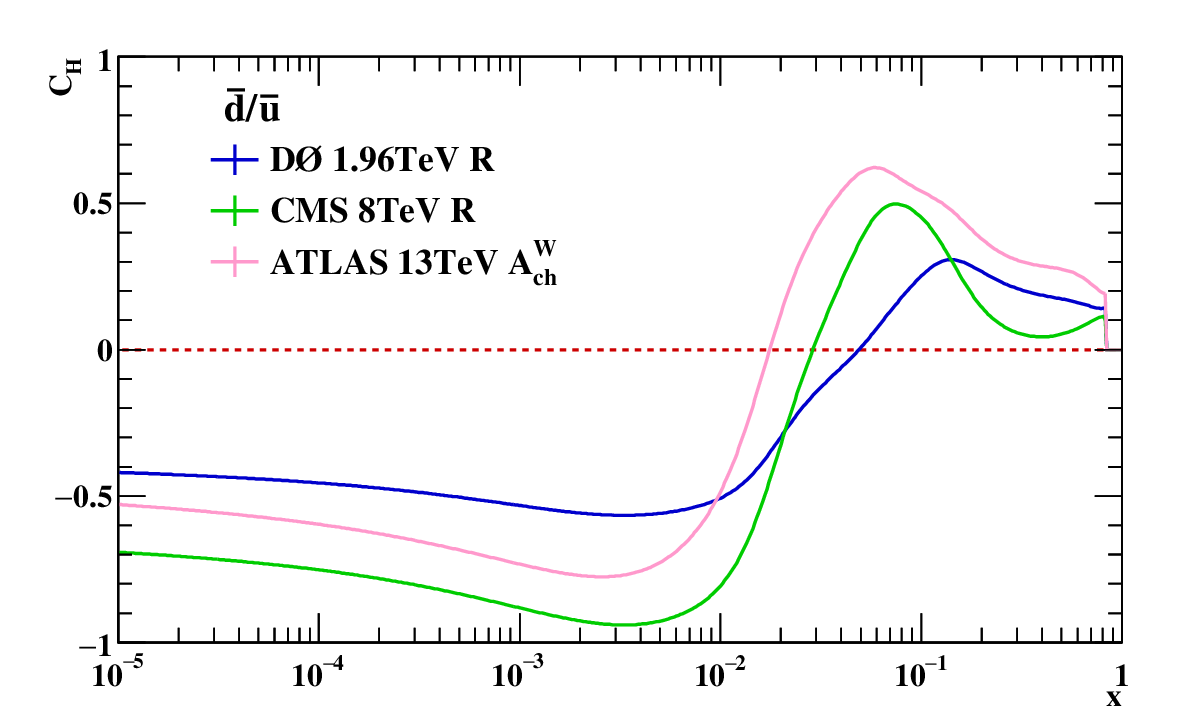}\\
	\caption{$C_H(E, f_\text{parton})$ (parton = $u_v$, $d_v$, $d_v/u_v$, $u$, $d$, $d/u$, $\bar{u}$, $\bar{d}$ and $\bar{d}/\bar{u}$, respectively) as a function of $x$, showing the correlation between each dataset and the parton PDF $f_\text{parton}$ from CT18NNLO.}
	\label{fig:CosH_Data}
\end{figure}

To identify which parton distribution functions are primarily responsible for the potential deviations observed among the three datasets, and to assess the consistency of their $x$-dependent correlation patterns, the \chEf between the new datasets and the $u_v$, $d_v$, $d_v/u_v$, $u$, $d$, $d/u$, $\bar{u}$, $\bar{d}$ and $\bar{d}/\bar{u}$ PDFs are shown in Figure~\ref{fig:CosH_Data}. The used quark PDFs are from CT18NNLO. The \Dzero 1.96 TeV $R$ measurement (the blue curve), the CMS 8 TeV $R$ measurement (the green curve) and the ATLAS 13 TeV $A^W_\text{ch}$ measurement (the pink curve) are plotted together. It should be emphasized that the exact definition of $R$, as given in Reference~\cite{CMS}, is employed in this chapter and in Chapter~\ref{sec:tension}, rather than the approximate expressions presented in Equations~\ref{equ:apr1} and \ref{equ:apr2}. 

Very similar patterns of behavior are observed among the different \chEf distributions in Figure~\ref{fig:CosH_Data}, reflecting the common physical sensitivity of the three new measurements. 
For the valence $u$ quark PDF ($u_v$, top-left panel), all three datasets yield positive \chEf values in the $x$ region around 0.1, corresponding to the larger-$x$ parton in their kinematics. From the definition of $C_H$, a positive correlation generally indicates that minimizing $\chisquare_E$ for a given dataset is accompanied by a reduction in the value of $f_{u_v}$ in this region. This consistent trend implies that all three measurements favor a lower $u_v$ compared with the CT18NNLO prediction at \xsim, in line with the observed excess of $R$ and the deficit in $A^W_\text{ch}$ discussed in Section~\ref{sec:data}. 
A complementary pattern is seen for the valence $d$ quark PDF ($d_v$, top-middle panel), where all datasets show negative \chEf values in the same $x$ region. This generally indicates that these datasets tend to enhance $d_v$ in this region, together with a reduction in $u_v$, implying that the deviations in $R$ and $A^W_{\text{ch}}$ can be jointly interpreted as a preference for a larger $d_v/u_v$ ratio than modern PDF predictions, in agreement with the negative value of the valence $d$-to-$u$ ratio ($d_v/u_v$, top-right panel) at \xsim.

The middle-row panels show the correlations with the $u$, $d$, and $d/u$ PDFs.
At \xsim, $u$ exhibits a positive correlation for all datasets, again indicating a preference for a downward shift in this range, while $d$ exhibits the opposite sign, indicating an enhanced $d$ quark density. These patterns persist for the antiquark PDFs ($\bar{u}$, $\bar{d}$, and $\bar{d}/\bar{u}$, bottom panels), where $\bar{d}$ follows the same trend as $u$, and $\bar{u}$ follows the same trend as $d$ at \xsim. Consequently, these behaviors lead to a negative value of $d/u$ and a positive value of $\bar{d}/\bar{u}$ at \xsim, thereby reinforcing the conclusion that the observed deviations in $R$ and $A^W_{\text{ch}}$ can be simultaneously accounted for by a shift in the valence quark ratio $d_v/u_v$, defined as $(d - \bar{d})/(u - \bar{u})$.

In summary, the strong correlation patterns for $u_v$, $d_v$, and $d_v/u_v$, together with the consistent behavior in the $u$, $d$, $d/u$, $\bar{u}$, $\bar{d}$, and $\bar{d}/\bar{u}$ PDFs, provide quantitative evidence that the three new datasets are pulling the current PDFs in the same direction of increasing $d_v/u_v$ ratio at \xsim. This mutual agreement supports the conclusion that the observed deviations are unlikely to be independent statistical fluctuations, but rather reflect a coherent deviation with the central predictions of current PDFs.

\section{Compatibility and Tension Analysis via Weighted PDF Updating}
\label{sec:tension}

To further assess the compatibility between the three new hadron collider measurements and the datasets currently input to modern global PDF fits, as well as to investigate their impact on the PDF updating results, the error PDF updating method package (\epump)~\cite{epump, epump2} is employed to update the CT18NNLO~\cite{CT18NNLO} PDFs. \epump~facilitates efficient updating of the best-fit PDF set and Hessian eigenvector pairs of PDF sets (i.e., error PDFs) in light of new data by retaining all theoretical assumptions of the original global fit, such as the choice of parametrization and number of parameters.
In Reference~\cite{epump2}, the \epump~framework is validated through detailed comparisons with full global analyses, and its potential is illustrated via selected phenomenological applications relevant to the LHC. 
In this work, the probed $x$ regions and the related energy scale $Q$ regions lie within the validated applicability range of Reference~\cite{epump2}. Therefore, despite the fixed assumptions, the qualitative conclusions regarding the impact of the new measurements remain valid.

The PDF updating is performed by systematically varying the statistical weights of the three newly input datasets ($R_\text{\Dzero}$, $R_\text{CMS}$, and $A^W_\text{ch ATLAS}$) from 0 to 15, thereby progressively enhancing their influences in the PDF updating process.
This procedure enables an assessment of how the global fit responds, and whether these measurements are consistent or in tension with the existing inputs.
Among the pre-existing datasets included in the modern PDFs, the NMC~\cite{nmc} and NuSea~\cite{nusea} data exhibit the most significant tensions with the three new hadron collider Drell--Yan measurements and are therefore highlighted. The NMC dataset provides values of the structure function ratio $F_2^D/F_2^p$ extracted from deep inelastic muon scattering on hydrogen ($p$) and deuterium ($D$), which is sensitive to the relative flavor composition of $u$ and $d$ quarks. 
The NuSea dataset is a Drell--Yan dimuon measurement with proton and deuteron targets, 
providing direct access to the $\bar{d}/\bar{u}$ ratio through the comparison of $pp$ and $pD$ cross sections, and reporting a pronounced enhancement of the $\bar{d}/\bar{u}$ distribution.

\begin{figure}[hbtp!]
	\centering
	\includegraphics[width=0.57\linewidth]{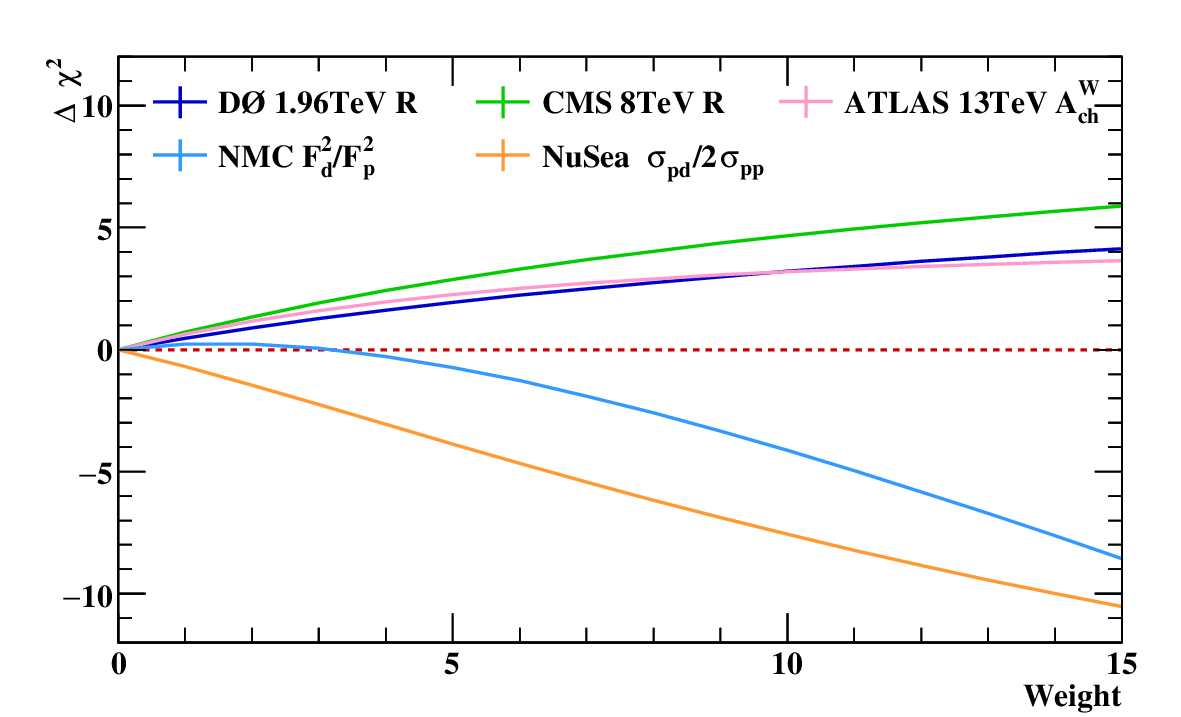}
	\caption{Impact on the global fit by increasing the weight (from 0 to 15) of the three new input datasets ($R_\text{\Dzero}$, $R_\text{CMS}$, $A^W_\text{ch ATLAS}$). 
		The effect is quantified by the changes in $\chisquare$ for the five selected datasets. Each curve shows $\Delta\chisquare = \chisquare(E)_\text{original PDF} - \chisquare(E)_\text{updated PDF}$ as a function of the updating weight applied to the corresponding dataset. An increasing slope reflects aligned PDF preferences, whereas a decreasing slope indicates opposite tendencies and potential tension.} 
	\label{fig:smart}
\end{figure}

As shown in Figure~\ref{fig:smart}, the shifts in the updating are quantified by evaluating the changes in $\chisquare$ for selected datasets, expressed as $\Delta\chisquare = \chisquare_\text{original PDF} - \chisquare_\text{updated PDF}$. In this representation, an increasing $\Delta\chisquare$ curve indicates that the selected dataset favors PDF modifications ‌in the same direction‌ as the three new data inputs, while a decreasing trend suggest a tension. The three new hadron collider Drell--Yan datasets exhibit monotonically increasing trends, indicating a consistent preference for the updated PDFs. In contrast, the NMC and NuSea datasets show generally decreasing curves with larger $\chisquare_\text{updated PDF}$ values, highlighting a significant tension between these fixed-target measurements and the new hadron collider Drell--Yan data.

\begin{figure}[htbp]
	\centering
	\includegraphics[width=0.32\textwidth]{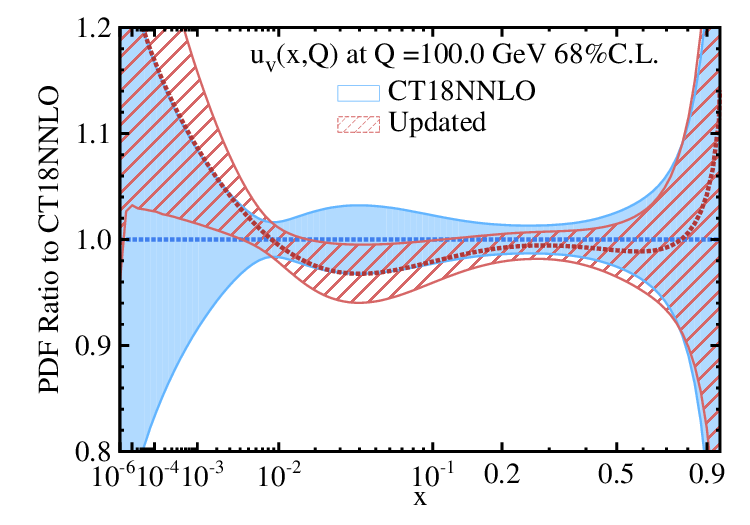}
	\includegraphics[width=0.32\textwidth]{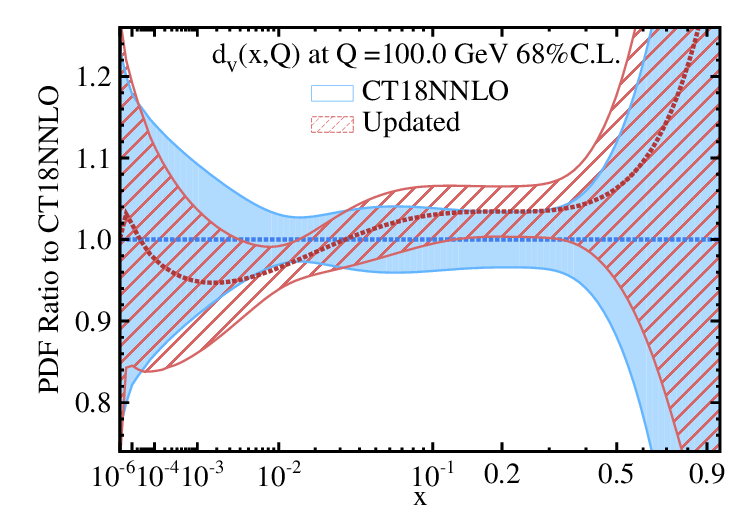}
	\includegraphics[width=0.32\textwidth]{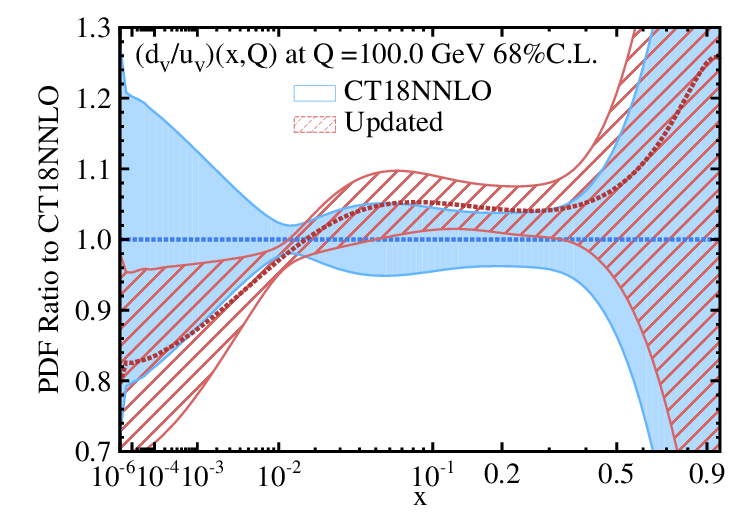} \\
	\includegraphics[width=0.32\textwidth]{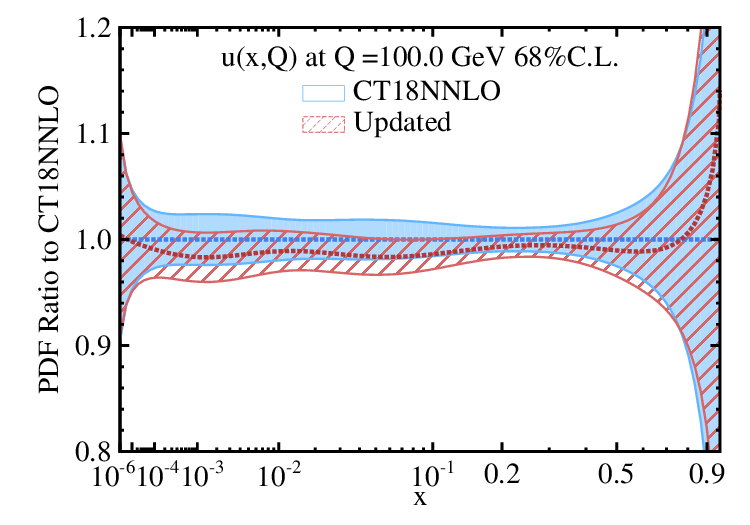} 
	\includegraphics[width=0.32\textwidth]{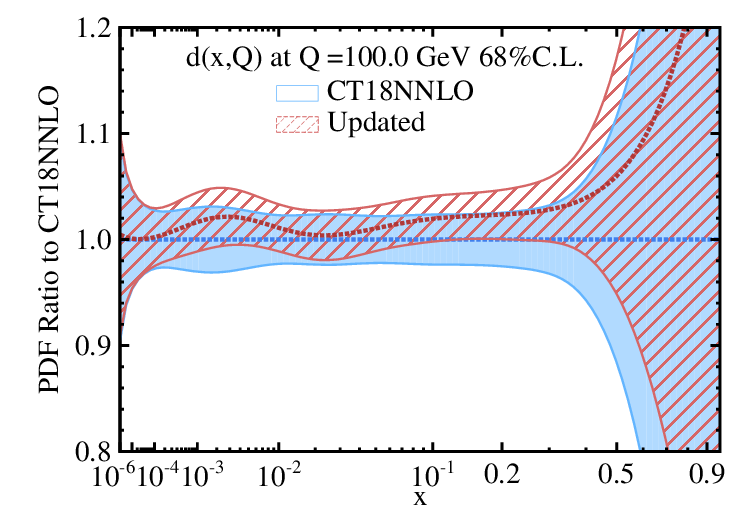} 
	\includegraphics[width=0.32\textwidth]{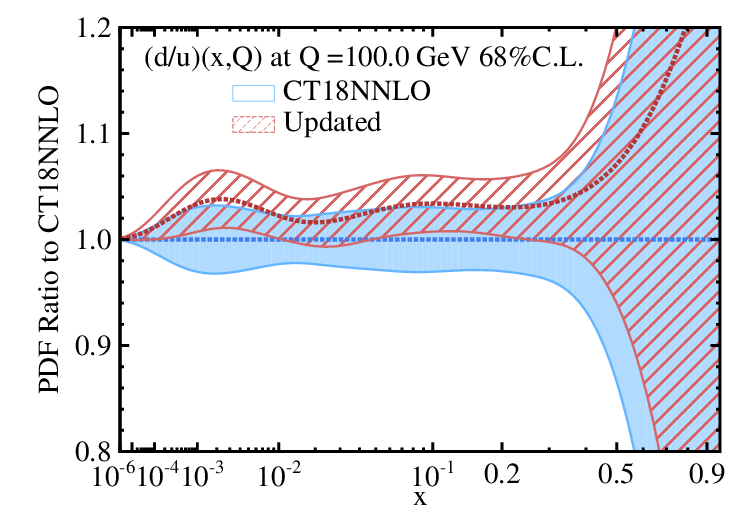} \\
	\includegraphics[width=0.32\textwidth]{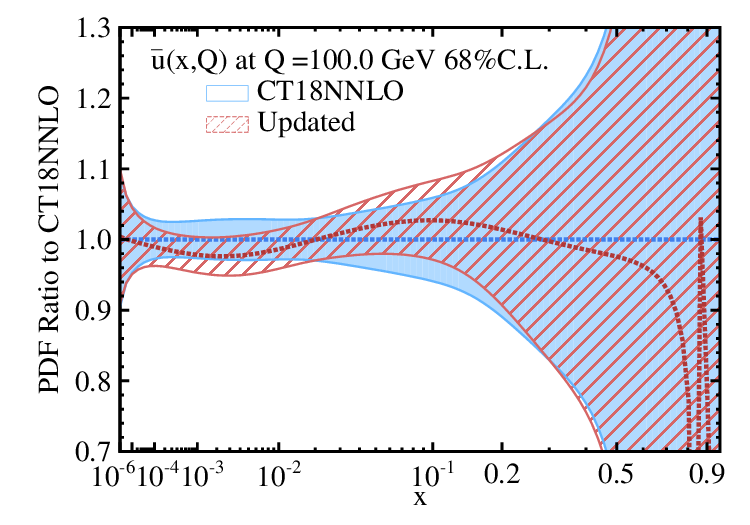} 
	\includegraphics[width=0.32\textwidth]{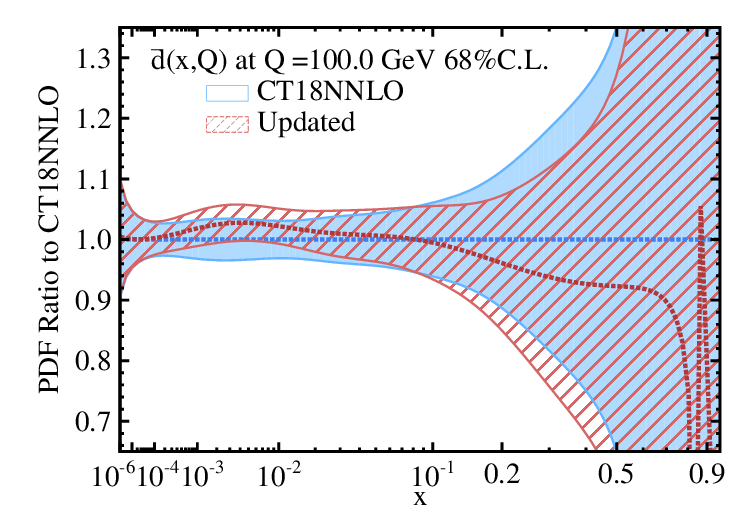}  \includegraphics[width=0.32\textwidth]{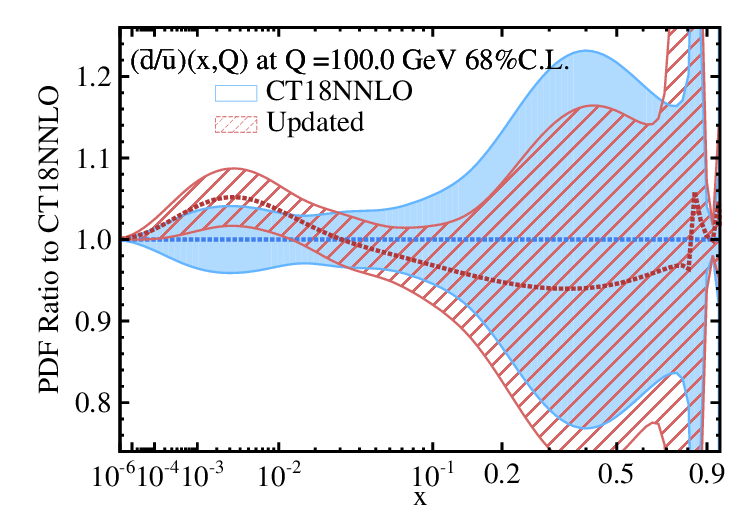} \\
	\caption{Comparison between updated PDFs, which are updated by the three new datasets with weight 15, and the original CT18NNLO central predictions. Each plot shows the ratio of the updated PDF to the original one, with associated uncertainty bands.}
	\label{fig:PDFUpdate}
\end{figure}

To further illustrate the impact of the three new datasets on individual parton densities, we compare the original CT18NNLO PDF set with the updated version generated through \epump~tool, where each dataset was assigned a weight of 15. Figure~\ref{fig:PDFUpdate} shows the ratios of the updated PDFs to the original ones, together with their uncertainties, for nine selected parton flavors: $u_v$, $d_v$, $d_v/u_v$, $u$, $d$, $d/u$, $\bar{u}$, $\bar{d}$ and $\bar{d}/\bar{u}$, all evaluated at $Q = 100$ GeV.

The most pronounced changes appear in the valence quark PDFs within the \xsim~region, where all three datasets have their strongest sensitivity. Consistent with the $C_H(E,f)$ correlation patterns in Figure~\ref{fig:CosH_Data}, the updated $u_v$ distribution exhibits a downward shift in this range, while $d_v$ shows a corresponding enhancement, jointly indicating a preference for a larger $d_v/u_v$ ratio. The same flavor-dependent tendency is also visible in the other quark PDFs: $u$, $\bar{d}$, and $\bar{d}/\bar{u}$ are reduced, whereas $d$, $\bar{u}$, and $d/u$ are enhanced, in agreement with the correlation analysis discussed in Section~\ref{sec:corr}. Outside the \xsim~region, the observed shifts are largely driven by the valence quark sum rules $\int_0^1 u_v(x)\text{d}x = 2$ and $\int_0^1 d_v(x) \text{d}x = 1$.

\section{Summary}
\label{sec:sum}

In summary, recent hadron collider Drell--Yan measurements including the $R$ ratios from forward-backward asymmetry at \Dzero and CMS, and the high-$m_T$ $W$ charge asymmetry $A^W_{\text{ch}}$ from ATLAS, exhibit a coherent preference for an enhanced $d_v/u_v$ ratio in the \xsim~region. 
Despite the differences in collider types, energies, and observables, these measurements demonstrate strong mutual consistency, especially in their correlation patterns within the global fit parameter space.
A clear tension is observed between these collider measurements and fixed-target data such as NMC and NuSea, which have traditionally played a dominant role in constraining the relative light quark flavor structure. This result indicates a distinct trend in the preferred flavor composition of $u$ and $d$ quarks. Future measurements with higher statistics will be crucial for placing more stringent constraints on the light-quark compositions. It should be noted that in this study, the results are derived from the \epump~analysis tool, and the associated momentum fraction $x$ represents only a Leading Order approximation, therefore a full PDF global analysis is ‌required for more robust conclusions.

\section{ACKNOWLEDGEMENTS}
\label{sec:ack}
This work was supported by the National Natural Science Foundation of China under Grant No. 12061141005 and 12105275.

\end{document}

%% file: author_list.tex
%
\affiliation{School of Physical Sciences, University of Science and Technology of China, Jinzhai Road 96, Hefei 230026, China}
%
\author{Zihan Zhao} \affiliation{School of Physical Sciences, University of Science and Technology of China, Jinzhai Road 96, Hefei 230026, China}
\author{Minghui Liu~\footnote{hepglmh@ustc.edu.cn}} \affiliation{School of Physical Sciences, University of Science and Technology of China, Jinzhai Road 96, Hefei 230026, China}
\author{Liang Han} \affiliation{School of Physical Sciences, University of Science and Technology of China, Jinzhai Road 96, Hefei 230026, China}
\vskip 0.25cm